\documentclass[11pt,twoside]{article}

\usepackage{asp2006}
\usepackage{graphicx}

\begin{document}
\title{What Do We Really Know about Mass Loss on the AGB?} 
\author{L. A. Willson}
\affil{Department of Physics and Astronomy, Iowa State University, 
Ames IA 50014}

\begin{abstract} 
Mass loss rate formulae are derived from 
observations or from suites of models.  For theoretical models, the 
following have all been identified as factors greatly influencing the 
atmospheric structure and mass loss rates:  Pulsation with piston 
amplitude scaling appropriately with stellar $L$;  dust nucleation 
and growth, with radiation pressure and grain-gas interactions and 
appropriate dependence on temperature and density; non-grey opacity 
with at least 51 frequency samples; non-LTE and departures from 
radiative equilibrium in the compressed and expanding flows; and 
non-equilibrium processes affecting the composition (grain formation; 
molecular chemistry).  No one set of models yet includes all the 
factors known to be important.  In fact, it is very difficult to 
construct a model that can simultaneously include these factors and 
be useful for computing spectra.  Therefore, although theoretical 
model grids are needed to separate the effects of $M,L,R$ and/or 
$T_{\mathrm{eff}} $or $Z$ on the mass loss rates, these models must be 
carefully checked against observations.  Getting the right order of 
magnitude for the mass loss rate is only the first step in such a 
comparison, and is not sufficient to determine whether the mass loss 
formula is correct.  However, there are observables that do test the 
validity of mass loss formulae as they depend directly on $d\log \dot 
M/d\log L$, $d\log \dot M/d\log R$, or $d\log \dot M/d\log P$.

\end{abstract}

\section{Reducing the Mass Loss Formulae to $ \dot M(L,M)$ along an 
Evolutionary Track}   

My analysis compares six published mass loss 
laws with the results of Bowen's 1995 model grid. Reimers' original 
formula (1975), was a fit of mass loss rate vs. $LR/M$ using an 
assortment of stars for which mass loss measurements were available 
in 1975; the same collection of observations could have equally well 
been fitted as functions of $R^{2}$ or other combinations of 
parameters (Goldberg 1979).  Vassiliadis and Wood (1993 = VW) fitted 
mass loss rates vs. period for pulsating AGB stars, mostly Miras. 
Van Loon et al. (2005) used data for supergiants.  Wachter et al. 
(2002) based theirs on models for carbon stars.  Schr{\"o}der and 
Cuntz (2005) proposed a correction to the Reimers' formula based on 
some theoretical arguments. Bl{\"o}cker (1995) proposed a 
modification to the Reimers' formula based on fitting models from 
Bowen (1988).  Finally, the Bowen and Willson (BW) prescription is 
based on a 1995 grid of models that incorporated some essential 
features not found in any of the others, or Bowen (1988):  Models 
followed the evolutionary tracks of individual stars, $R(L, M, 
\alpha, Z)$ where $\alpha$ is the mixing length parameter, and the 
driving amplitude of the "piston" boundary was regulated so that the 
peak power scales with stellar luminosity.  

Mass loss rates are 
sensitive to stellar parameters $L$, $R$ (or $P$ or $T_{\mathrm{eff}}$), and 
$M$.  To compare them, we use the definition of effective 
temperature, $L = 4\pi R^{2} \sigma T_{\mathrm{eff}}^{4}$; a 
period-mass-radius relation $P(M, R)$; and an evolutionary track 
$R(L, M,\alpha, Z)$.  This allows us to write all seven formulae as 
$\dot M(L, M)$ along a common set of evolutionary tracks.

\section{Comparison with Observational Constraints}

\subsection{The Death-Line and the Death Zone}

For each mass, 
$dM/dt=-(M/L_{D})dL/dt$ defines the deathline luminosity $L_{D}$; 
this yields critical mass loss rate $=(M/2\times 10^6)$ $M_{\odot} / 
yr$.  Four of the relations (Reimers, SC, VW, and BW) have ($M=1, 
\log L_{D} \approx 3.7 $) and ($M=2, \log L_{D} \approx 4.0$).  The 
van Loon et al. relation would be close to an extension of this death 
zone to higher masses, but its slope is too shallow to match the 
lower mass stars.   Wachter et al.'s carbon star mass loss rates have 
a relatively shallow slope and lower $L_{D}$, as does 
Bl{\"o}cker's.

Figure 1 shows mass loss rates for Reimers', VW, and BW formulae, and 
the deathline for each case.   It is clear that these two purely 
observational and quite independent empirical formulae agree 
substantially on the luminosity where the mass loss becomes 
important, although their $\dot M(L,M)$ are quite different.  I 
conclude that we know the deathline quite well, at least for stars 
between 1 and 2 solar masses.  The deathline satisfies $\dot M/M = 
\dot L/L$ and the death zone extends $\pm 1dex$ in mass loss rate 
above and below the deathline; the death zone's extent in $\log L$, 
and thus in time, is quite different for Reimers; formula compared 
with VW and BW.

\subsection{The Exponents}

Most tests of mass loss formulae have 
checked only whether they give the right order of magnitude of mass 
loss at the right range of stellar parameters.  However, two other 
tests are potentially much more constraining and informative:  Do 
they give the right $d\log \dot M /d\log L$ and the right $d\log \dot 
M/d\log M$?  It is in these slopes that the various relations differ 
most.  For shallow slopes, such as the Reimers', Wachter et al., and 
van Loon et al. examples, the mass loss occurs gradually as a star 
evolves up the AGB, so that by the time it reaches $L_{D}$ it has 
already lost a significant fraction of its original mass.  In 
constrast, the VW and BW formulae dictate that the mass is lost 
mostly near $L_{D}$.   Slopes $d\log \dot M/d\log L$ and $d\log \dot 
M/d\log M$  in the vicinity of $L_{D}$ for M = 1 and 2 are given for 
these seven formulae in Table 1, using on the Iben (1984) 
evolutionary tracks $R(L,M)$ and Ostlie and Cox (1986) $P(M,R)$.

\begin{table}[!hbt]
\begin{tabular}{l|lllllll}
\hline
    & R75  & VW & B & W & SC & vL & BW\\
\hline
   $d\log \dot M/d\log L$ for $M=1$ & 1.68 & 15.2 & 4.4 & 3.1 & 0.77 &
1.26 & 1.43  \\
   $d\log \dot M/d\log L$ for $M=2$ & 1 & 15.2 & 4.4 & 3.1 & 0.77 &
1.26 & 14.3  \\
   $d\log \dot M/d\log M$ for $M=1$ & 1 & $\sim$11 & 3.1 & 1.95 & 0
& 1.64 & 19.3  \\
   $d\log \dot M/d\log M$ for $M=2$ & 1.31 & $\sim$ 25 & 3.4 & 3.0 &
0.98 & 1.64 & 19.3  \\
\hline
\end{tabular}
\caption{Slopes from Seven Mass Loss Formulae:  Reimers (1975), 
Vassiliadis and Wood (1993), Bl{\"o}cker (1995), Wachter et al 
(2002), Schr{\"o}der and Cuntz (2005), van Loon et al (2005), and 
Bowen and Willson (see Willson 2000, 2006).}
\end{table}

The slopes may be compared directly with observational constraints. 
The slope $d\log \dot M/d\log L$ determines the amplitude of 
variation of the mass loss rate when the luminosity and/or radius of 
a star is varying, for example over a shell flash cycle.  For the 
shell flash variation we could, in principle, use $L(t)$ and $R(t)$ 
from evolutionary models, but these show that a very good 
approximation is obtained by using just $L(t)$ and assuming the shell 
flash variation stays on the same evolutionary track.  So $\Delta 
\log \dot M \approx (d\log \dot M/d\log L) \Delta \log L$; for a 
shell flash cycle $\Delta \log L \approx 0.4$.  Observations (e.g. 
Olofsson et al 1990; Sch{\"o}ier, Lindqvist, and Olofsson 2005; Decin 
et al. 2006) suggest that the variation is at least 1 or 2 dex, a 
result consistent only with the VW observational result and the BW 
models of the seven formulae tested.  There are significant secular 
and quasi-periodic variations in $P$ for about 10 \% of all Miras 
(Templeton et al 2005), an opportunity for further testing of the 
luminosity (or $R$ or $T_{\mathrm{eff}}$) dependence.

The slope $d\log \dot M/d\log M$ is harder to derive from 
observations, as it requires either that we know $M$ for individual 
stars with measured $\dot M$ or that we are able to fit a trend.  The 
scatter of observations around the VW fit, about $\pm$ 1dex, is an 
indication that $d\log \dot M/d\log L$ is large; if the range of 
masses included at a given $P$ is taken from evolutionary tracks 
crossing the Mira period-luminosity relation (VW Figure 20), then the 
scatter in mass loss rate around the VW fit suggests $d\log \dot 
M/d\log M \geq 8$; however, both the numerator and the denominator 
are sensitive to observational error.

An observational quantity related to $d\log \dot M/d\log M$ is the 
duration of the mass loss phase.  This may be defined as the time it 
takes the mass loss rate to increase by 2 dex near $L_{D}$, and we 
have used $\log L_{D} \pm$ 1 dex for the death zone.  For use with 
observational data, where $M$ and hence $L_{D}$ are not well known, 
the duration from $\log \dot M = $-7 to -5 will work equally well and 
give nearly the same answer.  When there is substantial positive 
feedback between mass loss and mass loss rates, this shortens the 
duration of the mass loss phase substantially:  For example, for the 
BW prescription with $d\log \dot M / d\log M$ artificially set to 
zero we get a duration (from $10^{-7}$ to $10^{-5} M_{\odot}/yr$) of 
$3\times 10^{5}$ years  while for the full BW prescription this is 
$1\times 10^{5}$ years.  Observations - for example, the width of the 
Mira $P-L$ relationship - suggest that the actual duration is no more 
than 200,000 years.  

Ultimately, the strongest test of mass loss 
relations near $L_{D})$ will be comparing the observed distribution 
of stars, $N(L)$, $N(P)$, or $N(T_{\mathrm{eff}})$ with those obtained from 
evolution models using a well-chosen mass loss formula.

\section{Conclusions}   

It is very difficult, perhaps impossible, 
to derive a mass loss prescription for use in stellar evolution 
directly from a set of mass loss observations, for the simple reason 
that measurable rates occur only for a narrow range of stellar 
parameters.  Observational fits of AGB stars agree what those stellar 
parameters are, giving consistent deathtlines $L_{D}(M)$.   To 
determine the dependence of the mass loss rates on stellar parameters 
for a given star as it evolves, we need to use models and indirect 
observational constraints.  Thus we can constrain $d\log \dot M/d\log 
L$ at a given $L$ using the amplitude of variation of the mass loss 
rate for a given $\Delta \log L$ or $\Delta \log P$, particularly 
near the deathline defined by $L_{D} = (M (dL/dt)/(dM/dt))$.  We can 
also constrain $d\log \dot M/d\log M$, for a given $d\log \dot 
M/d\log L$, by the duration of the mass loss phase, or the width of 
the death zone.  For currently available data, and using a standard 
set of evolutionary tracks, $d\log \dot M/d\log L \geq 14$; $d\log 
\dot M/d\log M$ is probably also greater than 10 and may be as large 
as 20.

For such large values of the exponents, their precise values only 
modify the shape of the corner in $\log M$ vs. $\log L$; the general 
pattern is the same.  The mass remains essentially constant until the 
star reaches the deathzone, and then the envelope is removed in a 
short time.  For population studies, all we really need to know is 
the deathline; a very good approximation for all mass loss formulae 
fitting the deduced sensitivity to $L$ and $M$ is that they evolve at 
constant mass to $L_{D}$ and then their envelope disappears.  The 
details of the mass loss formulae will still matter to those who wish 
to model the post-AGB phases or match the structure of the outflow as 
modulated by changes in $L$ and $M$ in the final stages, but for 
overall evolution of populations all that is needed is $L_{D}(M, Z)$, 
and this we know quite well, at least for $M$ $\sim$ 1 to 2 
$M_{\odot}$ and $Z$ $\sim$ solar.

\question{Busso} I would like to underline one more difficulty: when implementing an 
empirical formula into a code, you use your model
estimates for the parameters, for example T$_{\mathrm{eff}}$. Since T$_{\mathrm{eff}}$ 
is wrong in models due to lack of molecular opacities, you end up
in using something which is actually meaningless!
\answer{Willson} Yes, thank you. I prefer to use L, M.


\begin{thebibliography}{}
\bibitem[]{} Bloecker, T.\ 1995, \aap, 297, 727
\bibitem[]{} Bowen, G.~H., \& Willson, L.~A.\ 1991, \apjl, 375, L53
\bibitem[]{} Bowen, G.~H.\ 1988, \apj, 329, 299
\bibitem[]{} Bowen, G.~H.\ 1990, Numerical Modelling of Nonlinear Stellar Pulsations Problems and Prospects, 155
\bibitem[]{} Bowen, G.~H.\ 1990, New York Academy Sciences Annals, 617, 104
\bibitem[]{} Bowen, G.~H.\ 1992, Instabilities in Evolved Super- and Hypergiants, North-Holland, edited by Jager, C. de; Nieuwenhuijzen, H., 104
\bibitem[]{} Decin, L., Hony, S., de Koter, A., Justtanont, K., Tielens, A.~G.~G.~M., \& Waters, L.~B.~F.~M.\ 2006, \aap, 456, 549 
\bibitem[]{} Goldberg, L.\ 1979, \qjras, 20, 361
\bibitem[]{} H{\"o}fner, S., Gautschy-Loidl, R., Aringer, B., \& J{\o}rgensen, U.~G.\ 2003, \aap, 399, 589 
\bibitem[]{} Iben, I., Jr.\ 1984, \apj, 277, 333 
\bibitem[]{} Olofsson, H., Carlstrom, U., Eriksson, K., Gustafsson, B., \& Willson, L.~A.\ 1990, \aap, 230, L13 
\bibitem[]{} Ostlie, D.~A., \& Cox, A.~N.\ 1986, \apj, 311, 864
\bibitem[]{} Nowotny, W., Aringer, B., H{\"o}fner, S., Gautschy-Loidl, R., \& Windsteig, W.\ 2005, \aap, 437, 273
\bibitem[]{} Nowotny, W., Lebzelter, T., Hron, J., H{\"o}fner, S.\ 2005, \aap, 437, 285
\bibitem[]{} Reimers, D.\ 1975, Memoires of the Societe Royale des Sciences de Liege, 8, 369
\bibitem[]{} Sandin, C., H{\"o}fner, S.\ 2003, \aap, 404, 789 
\bibitem[]{} Sch{\"o}ier, F.~L., Lindqvist, M., \& Olofsson, H.\ 2005, \aap, 436, 633
\bibitem[]{} Schr{\"o}der, K.-P., \& Cuntz, M.\ 2005, \apjl, 630, L73
\bibitem[]{} Struck, C., Smith, D.~C., Willson, L.~A., Turner, G., \& Bowen, G.~H.\ 2004, \mnras, 353, 559
\bibitem[]{} Templeton, M.~R., Mattei, J.~A., \& Willson, L.~A.\ 2005, \aj, 130, 776
\bibitem[]{} van Loon, J.~T., Cioni, M.-R.~L., Zijlstra, A.~A., \& Loup, C.\ 2005, \aap, 438, 273
\bibitem[]{} Vassiliadis, E., \& Wood, P.~R.\ 1993, \apj, 413, 641
\bibitem[]{} Wachter, A., Schr{\"o}der, K.-P., Winters, J.~M., Arndt, T.~U., \& Sedlmayr, E.\ 2002, \aap, 384, 452
\bibitem[]{} Willson, L.~A., \& Kim, A.\ 2004, ASP Conf.~Ser.~313: Asymmetrical Planetary Nebulae III: Winds, Structure and the Thunderbird, 313, 394
\bibitem[]{} Willson, L.~A., Bowen, G.~H., \& Struck, C.\ 1996, ASP Conf.~Ser.~ 98: From Stars to Galaxies: the Impact of Stellar Physics on Galaxy Evolution, 98, 197
\bibitem[]{} Willson, L.~A.\ 2000, \araa, 38, 573
\bibitem[]{} Willson, L.~A.\ 2006, ESO Astrophysics Symposia, Springer, Planetary Nebulae Beyond the Milky Way,   99
\end{thebibliography}
\end{document}